\documentclass[prb,floatfix,twocolumn,showpacs,amsmath,amssymb]{revtex4}
\usepackage{graphicx}
\usepackage{dcolumn}
\usepackage{bm}
\begin{document}

\title{Mott transition in bosonic systems: Insights from the variational approach}
\author{Manuela Capello,$^{1}$ Federico Becca,$^{2,3}$ 
Michele Fabrizio,$^{2,3,4}$ and Sandro Sorella,$^{2,3}$}
\affiliation{
$^{1}$ Laboratoire de Physique Th\'eorique, Universit\'e Paul Sabatier, CNRS, 31400 Toulouse, France\\
$^{2}$  International School for Advanced Studies (SISSA),
I-34014 Trieste, Italy \\
$^{3}$ CNR-INFM-Democritos National Simulation Centre, Trieste, Italy. \\
$^{4}$ International Centre for Theoretical Physics (ICTP), P.O. Box 586, I-34014 Trieste, Italy
}

\date{\today} 

\begin{abstract}
We study the Mott transition occurring for bosonic Hubbard models in one, two, 
and three spatial dimensions, by means of a variational wave function 
benchmarked by Green's function Monte Carlo calculations. We show that 
a very accurate variational wave function, constructed by applying a 
long-range Jastrow factor to the non-interacting boson ground state, can
describe the superfluid-insulator transition in any dimensionality.
Moreover, by mapping the quantum averages over such a wave function into
the the partition function of a classical model, important insights into the 
insulating phase are uncovered. Finally, the evidence in favor of anomalous 
scenarios for the Mott transition in two dimensions are reported whenever 
additional long-range repulsive interactions are added to the Hamiltonian.
\end{abstract}


\maketitle

\section{Introduction}

The recent advances achieved on cold atoms trapped in optical lattices have 
generated an increasing interest in the condensed matter community, since they 
allow experimental realizations of simple lattice models.~\cite{zoller,greiner}
A great advantage of these systems is the possibility to have a direct control 
of the parameters, such as the width of the bands and the strength of the 
interactions, that can be manipulated by varying the depth of the optical 
potential.~\cite{zwerger} Therefore, cold atoms on optical lattices give the 
unique opportunity to make a close contact with theoretical models and to 
examine the origin of the fundamental physical phenomena that occur in 
crystalline materials. In particular, one of the most spectacular example is 
given by the superfluid-insulator transition in a system of interacting 
bosons: The so-called Mott transition.~\cite{greiner}

In this paper, we consider the superfluid-insulator transition in a bosonic 
system with different kinds of interactions. The simplest model that contains 
the basic ingredients of strong correlations is the Hubbard model
\begin{equation}\label{hambose}
{\cal H} = -\frac{t}{2} \sum_{\langle i,j \rangle} b^\dagger_i b_j + h.c. 
+ \frac{U}{2} \sum_i n_i (n_i-1),
\end{equation}
where $\langle \dots \rangle$ indicates nearest-neighbor sites, $b^\dagger_i$ 
($b_i$) creates (destroys) a boson on the site $i$, and $n_i=b^\dagger_i b_i$ 
is the local density operator. Here, we consider $N$ particles on 
a one-dimensional (1D) chain, a two-dimensional (2D) square lattice, and a 
three-dimensional (3D) cubic lattice with $L$ sites and periodic-boundary 
conditions. At zero temperature and integer densities $\rho=N/L$ there is a 
superfluid-insulator transition when the ratio between the kinetic energy and 
the on-site interaction is varied. Otherwise, for non-integer fillings, the 
ground state is always superfluid.
In a seminal paper,~\cite{fisher} by using a field-theoretical approach, 
Fisher and coworkers proposed that the transition of the $d$-dimensional 
clean system belongs to the $XY$ universality class in $d+1$. 
This scenario has been confirmed mostly in one and two dimensions by using 
different numerical techniques, such as quantum Monte Carlo and density-matrix 
renormalization group.~\cite{batrouni,batrouni2,kuhner,kuhner2,krauth}
In particular, it has been verified that in one dimension at $\rho=1$ 
there is a Kosterlitz-Thouless transition and the estimation of the critical 
value of the on-site interaction ranges between $U_c/t \sim 1.7$ and 
$U_c/t \sim 2.3$.~\cite{batrouni,kuhner}
Instead, a second-order phase transition is claimed to occur for 
$U_c/t \sim 8.5$ in two dimensions,~\cite{krauth} and for $U_c/t \sim 14.7$
in three dimensions.~\cite{prokofev}. Accurate estimations for the critical 
values can be also obtained from a strong-coupling 
expansion.~\cite{monien,monien2}

Besides the numerically exact techniques, important insights into the various
phases can be obtained by considering simplified variational wave functions.
The simplest example is given by the celebrated Gutzwiller state, where the
on-site correlation term allows for 
the suppression of the energetically expensive 
charge fluctuations. Contrary to the fermionic case, when considering 
bosons, it is possible to deal with this wave function without any further 
approximation.~\cite{kotliar,caffarel} Then, the Mott transition is obtained 
with a reasonable estimate of the corresponding critical value $U_c/t$, 
namely $U_c/t = d(\sqrt{n_c}+ \sqrt{n_c+1})^2$ for integer fillings 
$\rho=n_c$. The main drawback of this approach is that, similarly to what 
happens with fermions, the transition is reached with a vanishing kinetic 
energy and the insulating state completely lacks charge fluctuations, namely 
all particles are localized, $n_c$ in each lattice site.  Of course, this 
leads to a wrong description of the insulator, whenever the local interaction 
is finite. 

Following the ideas of previous works on fermionic 
systems,~\cite{capello,capello2,capello3} in a recent paper~\cite{capellobose}
we have shown that, in order to correct this outcome and obtain a more 
accurate description of the ground state, it is necessary to include a 
long-range Jastrow factor, whose singular behavior at small 
momenta was shown to be able to turn a non-interacting bosonic state into an 
insulator that still contains density fluctuations. In this paper, we present 
a more detailed study of the properties of that Jastrow wave function, as well 
as of its accuracy in comparison with Green's function Monte Carlo (GFMC) 
calculations, which, because of the absence of the sign problem, provide 
numerically exact results.~\cite{nandini,calandra} 

The same Jastrow wave function turns out to be also very effective to describe  
Hamiltonians that contain long-range interactions. This case has not 
been considered much in the literature, and we will show that different 
scenarios for the superfluid-insulator transition could occur. As a matter of 
fact, long-range interactions have been studied mostly in the continuum, where 
a transition between a charged bosonic fluid and a Wigner crystal has been 
found by varying the density.~\cite{magro,magro2,blatter}
In particular, in the presence of a logarithmic repulsion, the 2D Bose liquid
was found to have no condensate fraction, due to the predominance of 
long-wavelength plasmon excitations.~\cite{magro} In the last part of this 
paper, we generalize the Hubbard model of Eq.~(\ref{hambose}) to
\begin{equation}\label{hamboselr}
{\cal H}_{LR} = -\frac{t}{2} \sum_{\langle i,j \rangle} b^\dagger_i b_j + h.c. 
+ \frac{V}{2} \sum_{i,j} \Omega(R_i,R_j) n_i n_j,
\end{equation}
where $\Omega(R_i,R_j)$ is a long-range potential that only depends upon the 
relative distance $|R_i-R_j|$ between two particles and $V$ represents its 
strength. In particular, we will consider two possibilities for the long-range
potential. The first one is obtained by taking the Coulomb interaction between 
(charged) bosons moving in a 2D lattice embedded in a three-dimensional 
environment, which leads to the following 
potential in reciprocal space
\begin{equation}\label{Upoisson}
\Omega(q_x,q_y)=\frac{\pi}{\sqrt{(\cos q_x +\cos q_y -3)^2-1}},
\end{equation}
with a small-$q$ behavior $\Omega(q) \sim 1/|q|$. 
The second case consists in directly 
considering the solution of the Poisson equation in 2D:
\begin{equation}\label{Uq2}
\Omega(q_x,q_y)=\frac{1}{2-(\cos q_x +\cos q_y)},
\end{equation}
which for small momenta behaves like $\Omega(q) \sim 1/q^2$, leading to a 
logarithmic interaction in real space, i.e., $\Omega(r) \sim -\log (r)$.
In both cases, a uniform background is considered in order to cancel the 
divergent $q=0$ component of the potential.

The paper is organized as follow: In section~\ref{intro}, we describe the 
variational wave function, in sections~\ref{res1d},~\ref{res2d} 
and~\ref{res3d}, we show the results for the 1D, 2D and 3D short-range models. 
Then, in section~\ref{reslr}, we consider the 2D case with long-range 
interactions and finally, in section~\ref{conc}, we draw our conclusions.

\section{The variational wave function}\label{intro}

\subsection{The Jastrow wave function}\label{wavefunction}

The variational wave function is defined by applying a density Jastrow factor 
to a state with all the bosons condensed into the $q=0$ state
\begin{equation}\label{wfj}
|\Psi_J \rangle=
\exp \left \{ -\frac{1}{2} \sum_{i j} v_{i,j} (n_i-1)(n_j-1) \right \}
|\Phi_0 \rangle,
\end{equation}
where $|\Phi_0 \rangle=(\sum_i b^\dagger_i)^N |0 \rangle$ is the 
non-interacting boson ground state with $N$ particles, $\left(n_i-1\right)$ is 
the variation of the on-site density with respect to the average value 
$\rho=1$, and $v_{i,j}$ are translationally invariant parameters that can be 
optimized to minimize the variational energy.~\cite{sorella} 
In order to get some physical insight from the variational state, it is more 
instructive to consider the Fourier transform of the Jastrow parameters
$v_q$. Indeed, there is a tight connection between the small-$q$ behavior of 
$v_q$ and the nature of the ground state. In particular, as we are going to 
discuss, $v_q \sim 1/|q|$ implies the existence of sound modes in any 
dimensions, as expected in a superfluid. On the contrary, to recover an 
insulating behavior a much more singular $v_q$ for $q\to 0$ is 
required.~\cite{capellobose} The physical reason is that, in order to describe 
a realistic Mott insulating wave function that does include charge 
fluctuations, it is necessary to spatially correlate the latters. This is 
accomplished by a sufficiently singular $v_q$ that favors configurations where
opposite-sign fluctuations, $(n_i-1)(n_j-1)<0$, are close to each other,
while equal-sign ones, $(n_i-1)(n_j-1)>0$, are far apart.  

It should be mentioned that previous studies of fermionic 
systems~\cite{yokoyama} and more recent ones on the bosonic Hubbard 
model,~\cite{yokoyama2} stressed the importance of a many-body term
containing holon-doublon interactions for nearest-neighbor sites:
\begin{equation}\label{wfmb}
|\Psi_{g,MB} \rangle=
\exp \left ( -g \sum_i  n_i^2 + g_{MB} \sum_i \xi_i \right )|\Phi_0 \rangle,
\end{equation}
where $g$ and $g_{MB}$ are variational parameters and the many-body operator 
is defined by
\begin{equation}\label{manybody}
\xi_i=h_i \prod_{\delta} (1-d_{i+\delta})+d_i \prod_{\delta} (1-h_{i+\delta}),
\end{equation} 
where $h_i=1$ ($d_i=1$) if the site $i$ is empty (doubly occupied) and $0$ 
otherwise, $\delta=\pm x, \pm y$; therefore, $\xi_i$ counts the number of 
isolated holons (empty sites) and doublons (doubly occupied sites). 
Even though this operator has been originally introduced for fermionic 
systems, where the maximum occupancy at each site is given by two electrons,
it is useful also for bosons, since in the limit of large $U/t$ the number 
of sites with an occupation larger than two is negligible. However, within
this framework, the evidence for a true Mott transition is rather 
controversial, and it is not clear if an insulating phase can be stabilized
in the thermodynamic limit.~\cite{yokoyama2}

Combining together the variational Eqs.~(\ref{wfj}) and~(\ref{wfmb}), we
obtain the variational ansatz  
\begin{eqnarray}\label{wfjmb}
|\Psi_{J,MB} \rangle &=& 
\exp \left \{ -\frac{1}{2} \sum_{i,j} v_{i,j} (n_i-1)(n_j-1) \right . \nonumber \\
&& ~~~~~~~~+ \left .g_{MB} \sum_i \xi_i \right \} |\Phi_0 \rangle,
\end{eqnarray}
containing both the long-range Jastrow factor and a short-range many-body term.
As it will be shown in the next sections, the presence of the latter term is 
important in 2D and 3D, mainly to increase the accuracy in the 
strong-coupling regime. Instead, in 1D, the many-body term does not improve 
the accuracy of the long-range Jastrow state and there is no appreciable 
difference between the wave function~(\ref{wfj}) and~(\ref{wfjmb}).
In any case, we emphasize that, according to our calculations, the short-range
term alone cannot lead to an insulating behavior, and the main ingredient to
drive a superfluid-insulator transition is the long-range Jastrow factor,
parametrized by $v_q$. 

\subsection{Mapping onto a classical model}\label{Mapping}

The variational calculation with the wave function~(\ref{wfjmb}) can be shown 
to correspond to a classical problem at finite temperature. This correspondence
provides many insights into the properties of $|\Psi_{J,MB} \rangle$.  
To prove the mapping, let us denote a bosonic configuration by 
the positions $\{ x \}$ of the particles and a generic quantum state by 
$|\Psi \rangle$. For all the operators $\theta$ diagonal in space coordinates, 
the quantum average
\begin{equation}
\langle \theta \rangle = \frac{\langle \Psi | \theta |\Psi \rangle}
{\langle \Psi|\Psi \rangle}
\end{equation}
can be written in terms of the {\it classical} distribution
$|\Psi(x)|^2 = |\langle x|\Psi \rangle|^2 /
\sum_{x^\prime} |\langle x^\prime|\Psi \rangle|^2$, as
\begin{equation}
\langle \theta \rangle = \sum_x \langle x|\theta|x \rangle |\Psi(x)|^2.
\end{equation}
Since $|\Psi(x)|^2$ is a positive quantity, there is a precise correspondence 
between the wave function and an effective classical potential $V_{\rm cl}(x)$:
\begin{equation}\label{classical}
|\Psi(x)|^2 = e^{-V_{\rm cl}(x)}.
\end{equation}
The explicit form of the potential $V_{\rm cl}(x)$ depends upon the choice of 
the Jastrow factor, while $|\Phi_0 \rangle$ does not contribute to it, since
$\langle x|\Phi_0 \rangle$ does not depend upon the configuration $|x \rangle$.
In particular, whenever there is only the two-body potential (i.e., $g_{MB}=0$)
$V_{\rm cl}(x)$ is Gaussian, i.e., 
$V_{\rm cl}(x) = \sum_{q \ne 0} v_q n_{q}(x) n_{-q}(x)$, being $n_{q}(x)$ the 
Fourier transform of the local density of the configuration $|x \rangle$. 
In this case, the variational problem becomes equivalent to solve a classical 
model of oppositely charged particles (the holons and the doublons) mutually 
interacting through a potential determined by $v_q$. In the presence of the 
short-range many-body term, $V_{\rm cl}(x)$ is no longer Gaussian. However, 
when the density fluctuations are suppressed at large $U/t$, the quadratic 
term gives the most relevant contribution, hence the mapping onto a 
classical model of interacting oppositely charged particle still holds with  
$V_{\rm cl}(x) \sim \sum_{q \ne 0} v_q^{eff} n_{q}(x) n_{-q}(x)$, 
$v_q$ being replaced by a slightly different effective potential $v_q^{eff}$. 
In spite of the differences, it is plausible that the small-$q$ behavior of 
$v_q^{eff}$ must follow the same singular behavior of the Jastrow potential 
$v_q$. 

Let us now discuss in more details the connection between the form of $v_q$ and 
the low-energy excitation spectrum of the system. By means of the $f$-sum rule 
one can show that~\cite{roth,capellobose}
\begin{equation}
E_q = -2 \frac{\displaystyle \langle k\rangle}{\displaystyle D N_q} \sum_{i=1}^D\,\sin^2\left(\frac{q_i}{2}\right)= 
\frac{\displaystyle \int d\omega\,\omega\,N(q,\omega)}{\displaystyle \int d\omega\,N(q,\omega)},\label{E_q}
\end{equation}
where $N(q,\omega)$ is the dynamical structure factor, 
$N_q=\int d\omega\,N(q,\omega)$ the static one, and $\langle k \rangle$ the 
hopping energy per site. $E_q$ can be interpreted as the average excitation 
energy of density fluctuations at momentum $q$. $E_q\to 0$ for $q\to 0$ is a 
sufficient but not necessary condition for the existence of gapless 
excitations. In particular, deep inside the superfluid phase, $E_q$ must 
coincide with the energy dispersion of the Bogoliubov sound mode, which carries
most of the spectral weight. Although we can not access directly dynamical 
properties by our variational wave function, still we can provide a variational
estimate of $E_q$, in the same spirit of the Feynmann's construction in liquid 
Helium.~\cite{feynman} This amounts to use the variational values of 
$\langle k \rangle$ and of the static structure factor $N_q$, defined as
\begin{equation}
N_q= \frac{\langle \Psi| n_{q} n_{-q} |\Psi \rangle}
{\langle \Psi|\Psi \rangle},
\end{equation}
with $n_q$ the Fourier transform of the local density. Note that the 
uncorrelated  
\begin{equation}
N^0_q= \frac{\langle \Phi_0| n_{-q} n_{q} |\Phi_0 \rangle}
{\langle \Phi_0|\Phi_0 \rangle}
\end{equation}
is constant for any $q\not=0$ at $\rho=1$. Whenever a weak-coupling approach in
the Jastrow potential $v_q$ is possible, the following relation 
holds:~\cite{R&C}
\begin{equation}\label{nqvsvq}
N_q = \sum_x n_q(x)\,n_{-q}(x)\, e^{-V_{cl}(x)} \sim
\frac{N^0_q}{1+2 \, N^0_q v_q} \sim \frac{1}{v_q},
\end{equation}
the last equality following from the singular behavior of 
$v_q$ that is expected both in the superfluid and in the Mott insulating 
phases. Eq.~(\ref{nqvsvq}) shows that $v_q\sim 1/|q|$ allows to recover the 
correct behavior of $N_q\sim |q|$ and $E_q\sim |q|$ in the superfluid regime, 
which is what we indeed find, see following sections. In the insulating phase, 
should Eq.~(\ref{nqvsvq}) be valid, we would expect $v_q\sim \beta/q^2$ to get 
the expected behavior $N_q\sim q^2$ and $E_q$ finite for small $q$. 
This would correspond through Eq.~(\ref{classical}) to a classical Coulomb gas 
model (CGM) with effective temperature $T_{eff}=\pi/\beta$.~\cite{minnhagen}
In 1D, for any finite temperature, the CGM is always in a confined phase where 
oppositely charged particles are tightly bound in pairs, and with exponential 
decaying correlation functions.~\cite{lenard} This suggests that the 1D CGM may
indeed provide through the mapping~(\ref{classical}) a good description of a 
1D Mott insulating wave function.  

In 2D the CGM undergoes a Berezinskii-Kosterlitz-Thouless phase transition at 
$T_c^{CGM} \sim 1/4$, between a confined phase (stable at low temperature) and 
a plasma phase (stable at high temperature). Similarly to the 1D case, one 
would argue that the confined phase should correspond to the 2D Mott insulator. 
However, the 2D confined phase of the CGM displays power-law decaying 
correlations~\cite{minnhagen} that would correspond 
to power-law decaying equal-time correlations of the quantum ground state.  
This is not compatible with a genuine Mott insulator, which  already suggests 
that the insulating wave function must be characterized by a Jastrow potential 
$v_q$ more singular than $1/q^2$ as $q\to 0$, as indeed we find. 
In turns, this implies that (\ref{nqvsvq}) must not be valid, since, in spite 
of $v_q \times q^2\to \infty$ for $q\to 0$, we still expect $N_q\sim q^2$.    

The breakdown of Eq.~(\ref{nqvsvq}) becomes even more pronounced in 3D, where 
a potential $v_q \sim 1/q^2$ cannot describe at all an insulator since it
is not sufficiently singular to empty the condensate fraction.~\cite{reatto} 
In fact, we find that the optimized variational wave function shows a more 
diverging $v_q\sim 1/|q|^3$ in the 3D insulator,~\cite{capellobose} though 
$N_q\sim q^2$.  
The properties connected to the insulating phase can be again uncovered within 
a classical 3D gas with a $1/|q|^3$ potential, recently considered in 
Ref.~\onlinecite{3dcgm}. Indeed, analogously to what happens in 2D, also in 3D 
a system of charges interacting with a logarithmic potential in real space 
admits a high-temperature fluid regime separated by a low-temperature 
dielectric phase. Within this mapping, the 3D insulating state is found to 
correspond to the low-temperature phase of this classical model.

In the following, we will present our results obtained by considering the 
quantum variational wave function and we will use the classical mapping to 
gain insights into the ground-state properties. Moreover, in order to verify 
the accuracy of the variational calculations, we will perform the numerically 
exact GFMC that allows one to obtain ground-state properties.

\section{The 1D Hubbard model}\label{res1d}

\begin{figure}
\includegraphics[width=\columnwidth]{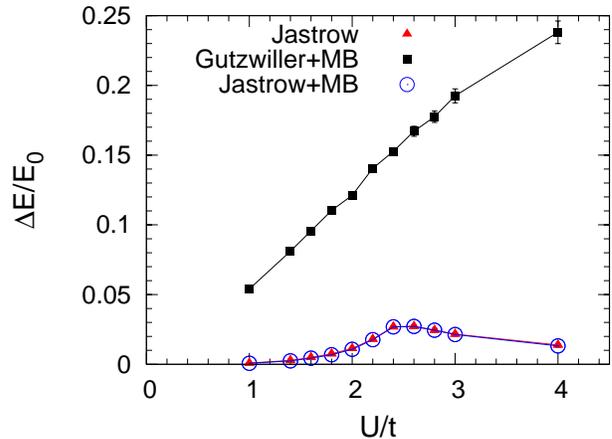}
\caption{\label{fig:accuracy1d} 
Accuracy of different variational wave functions as a function of $U/t$ for 
$60$ sites and $60$ bosons. $\Delta E= E_0 - E_{VMC}$, where $E_{VMC}$ is the
variational energy and $E_0$ is the ground-state one, obtained by GFMC.
The state of Eq.~(\ref{wfj}) is denoted by ``Jastrow'', the one of 
Eq.~(\ref{wfmb}) by ``Gutzwiller+MB'', and the one of Eq.~(\ref{wfjmb}) by
``Jastrow+MB''.}
\end{figure}

Here, we consider a chain of $L$ sites with periodic boundary conditions and 
$N=L$ bosons. First of all, in Fig.~\ref{fig:accuracy1d} we compare the 
variational accuracy of the wave functions~(\ref{wfj}),~(\ref{wfmb}) 
and~(\ref{wfjmb}) for different values of $U/t$. Once a long-range Jastrow 
factor is considered, the many-body term of Eq.~(\ref{manybody}) is irrelevant 
and there is no appreciable difference between the wave functions~(\ref{wfj}) 
and~(\ref{wfjmb}), for all the values of the on-site interaction. By contrast, 
the Gutzwiller state, also when supplied by the many-body term, is much less 
accurate by increasing $U/t$.

\begin{figure}
\includegraphics[width=\columnwidth]{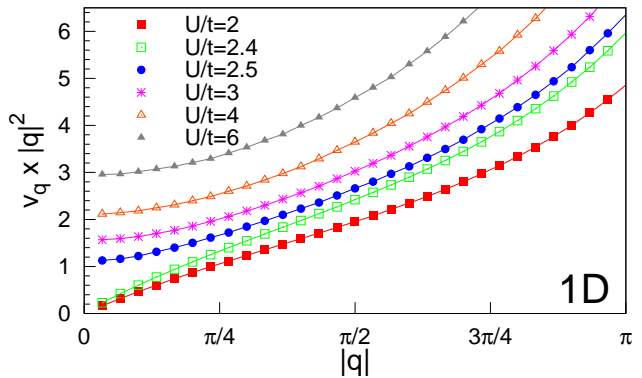}
\caption{\label{fig:jas1d} 
Jastrow parameters $v_q$ multiplied by $q^2$ as a function of $|q|$ for $60$ 
sites and $60$ bosons.}
\end{figure}

\begin{figure}
\includegraphics[width=\columnwidth]{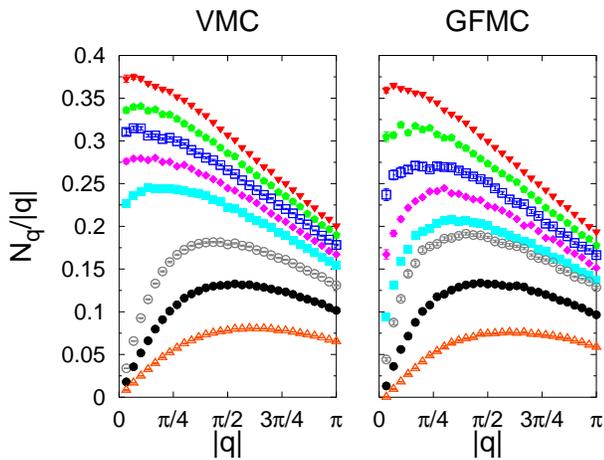}
\caption{\label{fig:nq1d} 
Density structure factor $N_q$ divided by $|q|$ calculated with variational 
Monte Carlo (left panel) and GFMC (right panel) for different $U/t$ and $L=60$. 
From top to bottom $U/t=1.6$, $1.8$, $2$, $2.2$, $2.4$, $2.5$, $3$, and $4$.}
\end{figure}

Therefore, in the following, we will consider the state with long-range
Jastrow factor~(\ref{wfj}) alone, since the many-body term makes the algorithm 
much slower and does not improve the quality of the variational state. 
In Fig~\ref{fig:jas1d}, we report the minimized Jastrow parameters $v_q$ 
multiplied by $q^2$ for different $U/t$: There is a clear difference in the 
small-$q$ behavior for $U/t \lesssim 2.4$, where $v_q \sim \alpha/|q|$ and 
for $U/t \gtrsim 2.5$, where $v_q \sim \beta/q^2$.
At the variational level, the change of the singular behavior of the Jastrow 
parameters for $U/t \sim 2.45$ marks the superfluid-insulator transition. 
Indeed, as discussed in the previous paragraph, $v_q \sim \alpha/|q|$ implies
a gapless system, whereas $v_q \sim \beta/q^2$ indicates a finite gap in the
excitation spectrum. Let us now concentrate on the insulating phase. Here, 
the Jastrow wave function~(\ref{wfj}) can be mapped onto the partition function
of an effective classical CGM and $\beta$ plays the role of the inverse 
classical temperature $\beta=\pi/T_{eff}$. 
In 1D the CGM is in the confined phase for any finite temperature, with 
exponential correlations.~\cite{lenard} This outcome is consistent with the
fact of having, in the quantum model, a finite gap in the excitation spectrum.
Remarkably close to the Mott transition in the insulating 
phase, the value of $\beta$ obtained from the optimized Jastrow potential is 
very small and approaches to zero when $U \to U_c$ from above. Since the
correlation length of the 1D CGM diverges for $\beta \to 0$, our numerical
outcome gives a strong indication in favor of a continuous transition between 
the superfluid and the Mott insulating phase.

Let us now analyze the transition by considering the density structure factor
$N_q$. In the small-$q$ regime, we can generally write that
\begin{equation}
N_q=\gamma_1 |q| + \gamma_2 q^2 +O(q^3),
\end{equation}
where $\gamma_1$ and $\gamma_2$ depend upon the Jastrow parameters.
In analogy with spin systems, we have that $\gamma_1=v_c\chi$, with $v_c$ and 
$\chi$ the sound velocity and the compressibility, respectively. The fact of 
having $\gamma_1=0$ in the insulating regime indicates that this state is 
incompressible (i.e., $\chi=0$). From Fig.~\ref{fig:nq1d}, we obtain that 
$\gamma_1$ has a very sharp crossover from a finite value to zero across the 
transition, suggestive of a true jump in the thermodynamic limit. This outcome 
is consistent with the fact that the compressibility also has a finite jump in 
the 1D quantum phase transition.~\cite{laflorencie} Moreover, just above 
$U_c$ in the insulating regime, $\gamma_2$ is very large for both the 
variational and the GFMC calculations, indicating the peculiar character of 
the 1D transition. By using the small-$q$ behavior of the density structure 
factor, the Mott transition can be located at $U_c/t \sim 2.45\pm 0.05$ for the
variational calculation, whereas the GFMC approach gives 
$U_c/t \sim 2.1\pm 0.1$.

\begin{figure}
\includegraphics[width=\columnwidth]{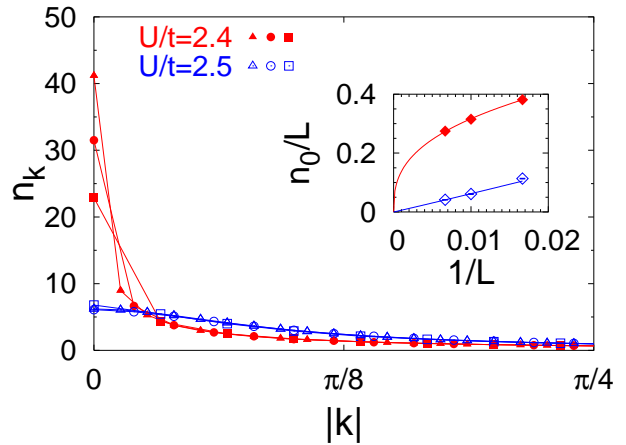}
\caption{\label{fig:nk1d} 
Variational results for the momentum distribution ${\tt n}_k$ in 1D for $L=60$
(squares), $100$ (circles), and $150$ (triangles) across the transition 
($U/t=2.4$ and $2.5$). Inset: Size scaling of the condensate fraction 
${\tt n}_0/L$ for $U/t=2.4$ (upper curve) and $U/t=2.5$ (lower curve).}
\end{figure}

The superfluid-insulator transition can be also easily detected by considering 
the momentum distribution:
\begin{equation}
{\tt n}_k=\frac{\langle\Psi_J| b^\dagger_k b_k|\Psi_J \rangle}
{\langle\Psi_J|\Psi_J \rangle},
\end{equation} 
where $b^\dagger_k$ is the creation operator of a boson of momentum $k$.
This quantity has a radically different behavior below and above the 
transition: In the superfluid phase, it has a cusp at $k=0$, although there
is no condensate fraction, i.e., ${\tt n}_0/L \to 0$ for $L \to \infty$, 
while in the insulating phase it is a smooth function of the momentum $k$, 
see Fig.~\ref{fig:nk1d}.

\begin{figure}
\includegraphics[width=\columnwidth]{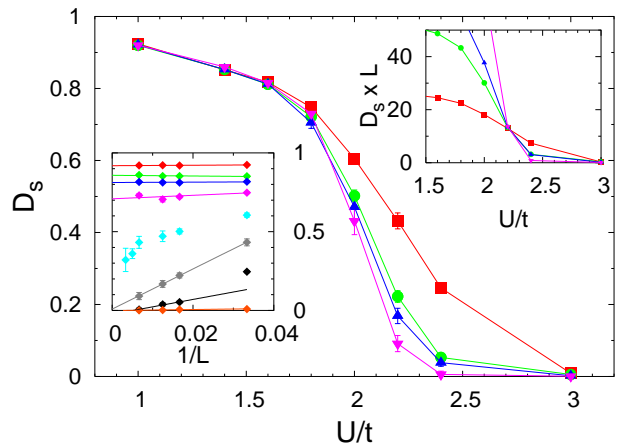}
\caption{\label{fig:drude1d} 
Superfluid stiffness $D_s$ calculated by GFMC as a function of $U/t$ for 
different sizes in the 1D boson Hubbard model. Lower inset: Size scaling of 
$D_s$ for different $U/t$ (same values of the main panel, with $U/t$ increasing
from top to bottom). Upper inset: $D_s \times L$ as a function of $U/t$. 
The point where the different curves cross marks the transition point.}
\end{figure}

Finally, we want to conclude the 1D part by considering the superfluid 
stiffness $D_s$. In analogy to what has been done by Pollock and Ceperley
at finite temperature,~\cite{pollock} this quantity can be also calculated 
directly at zero temperature by using the GFMC and the so-called winding 
numbers (see Appendix)
\begin{equation}\label{stiffness}
D_s=\lim_{\tau\to \infty} 
\frac{\langle \Psi_0 ||{\vec W}(\tau)|^2 |\Psi_0 \rangle}{D \; L \; \tau},
\end{equation}
where $|\Psi_0 \rangle$ is the ground-state wave function obtained by GFMC,
$D$ is the dimension of the system and 
${\vec W}(\tau)=\sum_i[{\vec r}_i(\tau)-{\vec r}_i(0)]$, ${\vec r}_i(\tau)$ 
being the position of the $i$-th particle after evolving it for a diffusion 
time $\tau$ from the initial position ${\vec r}_i(0)$. The diffusion process 
must be done without considering periodic boundary conditions, namely by 
increasing or decreasing the values of the coordinates of a particle that 
crosses the boundaries of the cluster. In this way, non-zero winding numbers 
across the lattice can be detected. It should be stressed that, exactly at zero 
temperature, $D_s$ can only give information about the presence of a gap in 
the excitation spectrum, and, therefore, it can discriminate between conducting
and insulating phases. Our results show that $D_s$ is finite and large in the 
weak-coupling regime, whereas it vanishes for $U/t \gtrsim 2.1$, 
see Fig.~\ref{fig:drude1d}. Again, in analogy with spin systems and from 
general scaling arguments valid for 1D boson models, we expect a jump 
at the transition, that however is very hard to detect by considering finite
clusters.~\cite{laflorencie} Nevertheless, an accurate value of the 
transition point is obtained from the size scaling of $D_s \times L$, 
see Fig.~\ref{fig:drude1d}.

\section{The 2D Hubbard model}\label{res2d}

\begin{figure}
\includegraphics[width=\columnwidth]{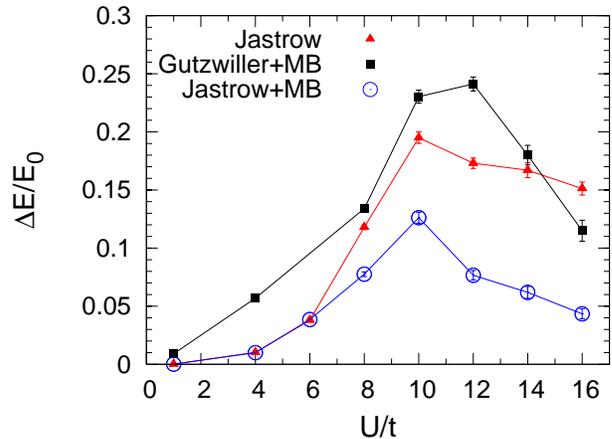}
\caption{\label{fig:accuracy2d} 
Accuracy of different variational wave functions as function of $U/t$ for 
the $10 \times 10$ cluster and $100$ bosons. The symbols are the same as in 
Fig.~\ref{fig:accuracy1d}.} 
\end{figure}

Let us now turn to the 2D Hubbard model and consider square clusters with 
$L=l \times l$ sites and $N=L$ bosons. The accuracy of the three wave 
functions~(\ref{wfj}),~(\ref{wfmb}) and~(\ref{wfjmb}) are reported in 
Fig.~\ref{fig:accuracy2d}. The situation is different from the previous 1D 
case: The Gutzwiller state with the many-body term, which in 1D is not accurate
for large $U/t$, in 2D becomes competitive with the Jastrow wave function 
for $U/t \gtrsim 14$. Moreover, the presence of the many-body term strongly 
improves the accuracy of the Jastrow state as soon as $U/t \gtrsim 8$. 
Then, in the following, we will consider the state~(\ref{wfjmb}) for both the 
variational and the GFMC calculations. 

\begin{figure}
\includegraphics[width=\columnwidth]{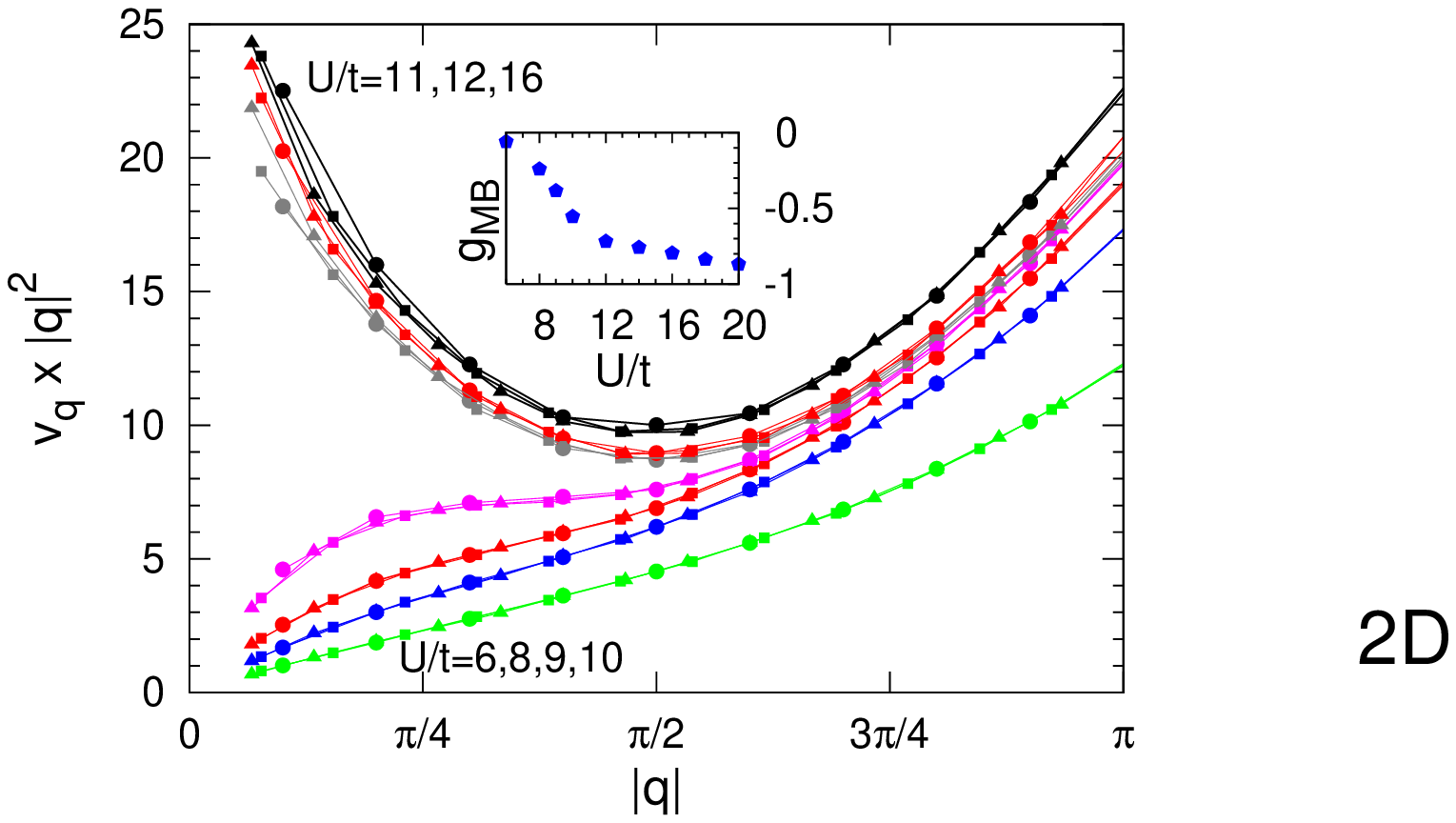}
\caption{\label{fig:jas2d} 
Jastrow parameters $v_q$ multiplied by $q^2$ as a function of 
$|q|$ [along the (1,0) direction] for $20 \times 20$ (circles), $26 \times 26$ 
(squares), and $30 \times 30$ (triangles) clusters. Inset: The many-body 
variational parameter $g_{MB}$ as a function of $U/t$.}
\end{figure}

\begin{figure}
\includegraphics[width=\columnwidth]{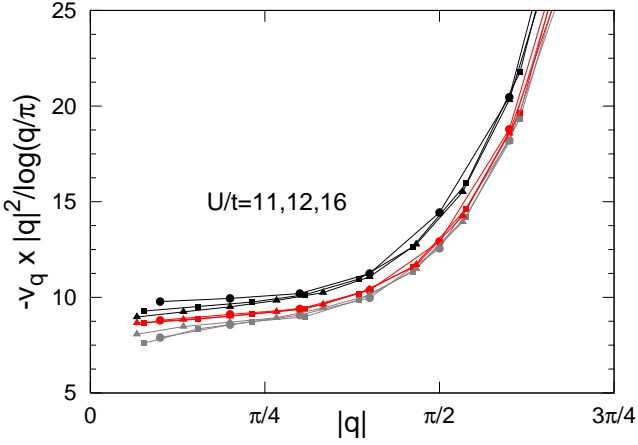}
\caption{\label{fig:jas2dlog}
Jastrow parameters $v_q$ multiplied by $q^2/\log(q/\pi)$ as a function of 
$|q|$ [along the (1,0) direction] for different $U/t$ and the same sizes as 
Fig.~\ref{fig:jas2d}.}
\end{figure}

\begin{figure}
\includegraphics[width=\columnwidth]{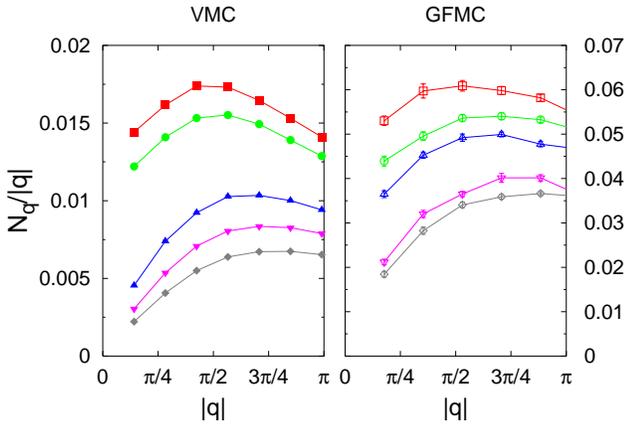}
\caption{\label{fig:nq2d} 
Left panel: Density structure factor $N_q$ divided by $|q|$ calculated by the 
variational Monte Carlo for different $U/t$ and $L=20 \times 20$. 
From top to bottom $U/t = 10$, $10.2$, $10.4$, $11$, and $12$.
Right panel: The same for the GFMC on the $L=16 \times 16$ cluster.
From top to bottom $U/t = 8$, $8.2$, $8.4$, $8.6$, and $8.8$.}
\end{figure}

In Fig.~\ref{fig:jas2d}, we show the behavior of the optimized $v_q$ as a 
function of the interaction strength. Similarly to what happens in the 1D 
system, we obtain that $v_q \sim \alpha/|q|$ for $U/t \lesssim 10.5$, while 
$v_q$ is best fitted by $v_q\sim -\log(q)/q^2$ for $U/t \gtrsim 10.5$
(see Fig.~\ref{fig:jas2dlog}), 
corresponding to the superfluid and the Mott insulating phase, respectively. 
As we anticipated in section~\ref{Mapping}, the Jastrow potential in the 
insulating phase is more singular than $1/q^2$, suggestive of a classical 
model with bound charges but presumably without the power-law correlations 
displayed by the CGM in the confined phase. In reality, on the sizes available 
to our numerical approach, we can not firmly establish whether a classical 
potential $-\log(q)/q^2$ has indeed exponential decaying correlation functions.
Nevertheless, it is remarkable and very encouraging that the variational 
optimization leads to such a singular $v_q$. 

\begin{figure}
\includegraphics[width=\columnwidth]{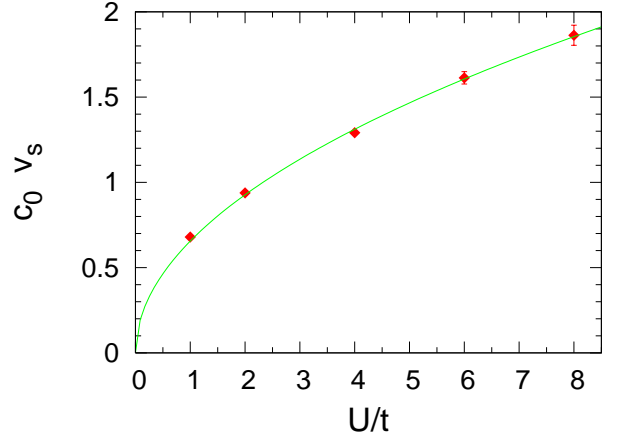}
\caption{\label{fig:vs} 
GFMC results for the sound velocity $v_c$ obtained through a finite
size scaling of the ground-state energy, see Eq.~(\ref{vsscaling}). The line
is a guide to the eyes.}
\end{figure}

\begin{figure}
\includegraphics[width=\columnwidth]{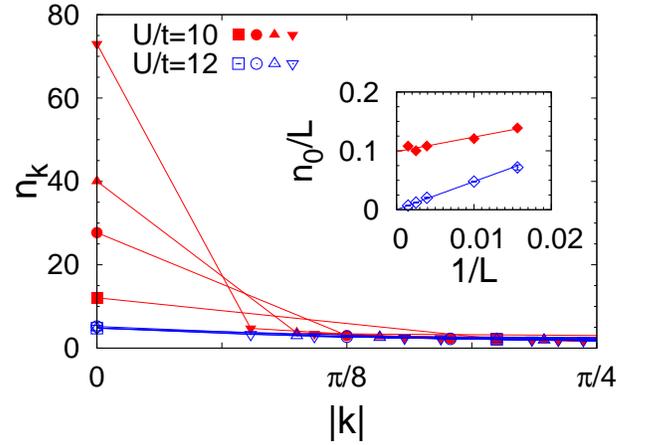}
\caption{\label{fig:nk2d} 
Variational results for the momentum distribution ${\tt n}_k$ in 2D for 
$10 \times 10$ (squares), $16 \times 16$ (circles), $20 \times 20$ 
(upward triangles), and $26 \times 26$ (downward triangles) clusters with 
$U/t=10$ and $12$. Inset: Size scaling of the condensate fraction 
${\tt n}_0/L$ for $U/t=10$ (upper curve) and $U/t=12$ (lower curve).}
\end{figure}

\begin{figure}
\includegraphics[width=\columnwidth]{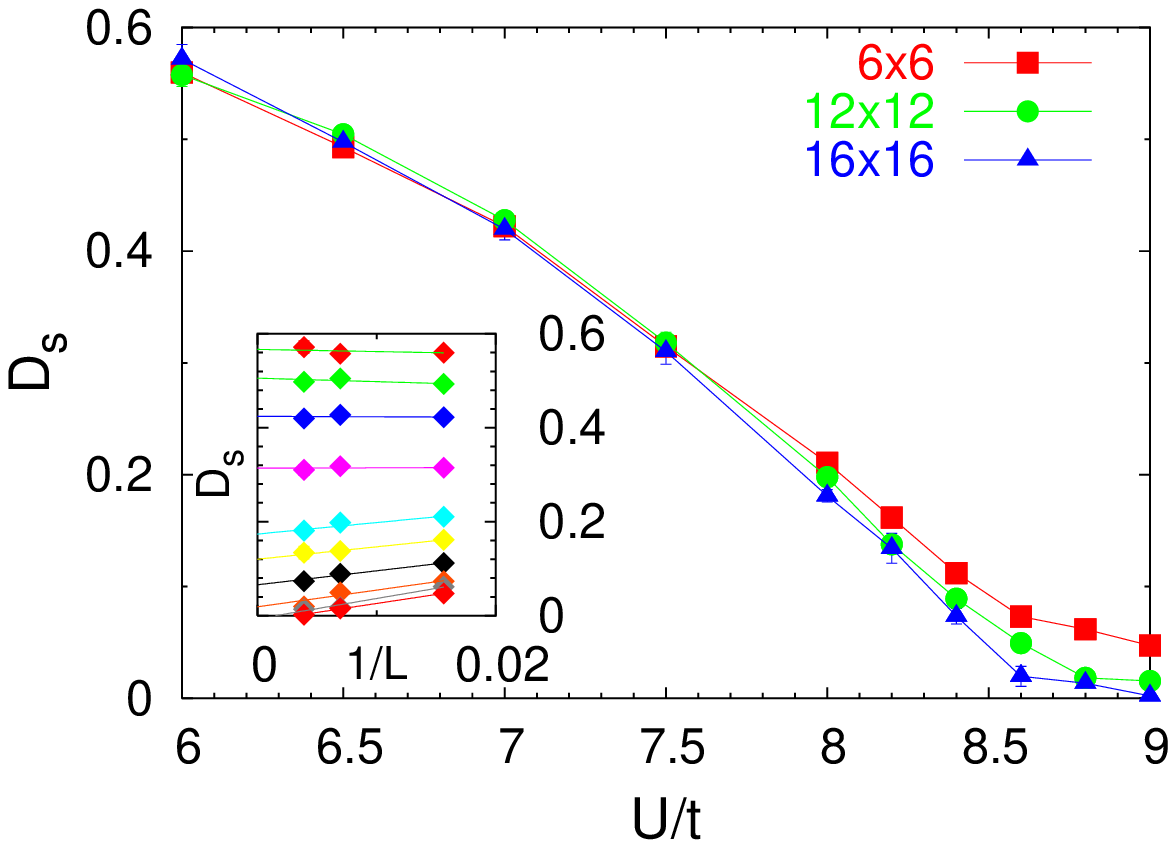}
\caption{\label{fig:drude2d} 
Superfluid stiffness $D_s$ as a function of $U/t$ for different sizes. 
Inset: Size scaling of $D_s$ for different $U/t$ (the same values of the main 
panel are reported, with $U/t$ increasing from top to bottom).}
\end{figure}

The evidence that the change in behavior of $v_q$ does correspond to the 
superfluid-insulator transition comes also by the small-$q$ limit of the 
structure factor $N_q$. In Fig.~\ref{fig:nq2d}, we show the results for the 
variational and the GFMC calculations as a function of $U/t$. In both cases, 
we find a different small-$q$ behavior for large and small couplings. In the 
variational calculations, for $U/t \lesssim 10.3$ the structure factor behaves 
as $N_q \sim \gamma_1 |q|$, while for $U/t \gtrsim 10.3$ we get 
$N_q \sim \gamma_2 q^2$. Therefore, at the variational level, the 
superfluid-insulator transition is located around $U_c/t \sim 10.3\pm 0.1$; 
this value is slightly smaller than the one obtained by the simple Gutzwiller 
wave function, for which $U_c/t \sim 11.65$.~\cite{caffarel} The critical 
value of the on-site interaction is rather different within GFMC, yielding 
$U_c/t \sim 8.5\pm 0.1$, in close agreement with the value found in the 
literature.~\cite{krauth} Let us focus more deeply on the behavior of the 
structure factor $N_q$. Approaching the transition from the weak-coupling 
region, $\gamma_1$ goes smoothly to zero, in contrast with the results of the 
1D model, where we observed an abrupt jump. Also contrasting the 1D case, 
$\gamma_2$ is found to be finite and continuous across the transition.  
These results, based upon the variational wave function, are 
confirmed by the GFMC calculations, see Fig.~\ref{fig:nq2d}.
The vanishing linear coefficient of $N_q$ can be ascribed either to $v_s$ or 
to $\chi$. In order to clarify which one of these quantities goes to zero at 
the transition, we extract the sound velocity $v_s$ from the finite-size 
scaling of the exact ground-state energy by
\begin{equation}\label{vsscaling}
\epsilon_0(L) = \epsilon_0(\infty) - \frac{c_0 v_s}{l^3},
\end{equation}
where $\epsilon_0(L)$ is the ground-state energy per site for a cluster with 
$L=l^2$ sites, $\epsilon_0(\infty)$ is the extrapolated value in the 
thermodynamic limit, and $c_0$ is a given model-dependent constant. 
Our results, shown in Fig.~\ref{fig:vs}, clearly indicate that $v_s$ stays 
finite across the superfluid-insulator transition, thus implying a vanishing 
compressibility when approaching the insulator. 

The fingerprint of this transition is also given by the momentum distribution,
see Fig.~\ref{fig:nk2d}. For this quantity, a striking difference is observed 
below and above $U_c$. In the former case, a cusp-like behavior with a finite 
condensate fraction is found, whereas in the latter case a smooth behavior is 
detected, with a vanishing ${\tt n}_0/L$. Notice that a vanishing condensate 
fraction in the thermodynamic limit immediately follows from $N_q \sim q^2$ 
by using the f-sum rule derived in Ref.~\onlinecite{stringari}.

Finally, we report in Fig.~\ref{fig:drude2d} the stiffness $D_s$, calculated
by using GFMC. In 2D, we obtain a different behavior with respect to the 1D 
case, where a finite jump is rather clear at the superfluid-insulator 
transition. Indeed, the evaluation of the stiffness for different sizes 
confirms the absence of the jump in 2D, as expected for a second order phase  
transition.~\cite{krauth}

\section{The 3D Hubbard model}\label{res3d}

\begin{figure}
\includegraphics[width=\columnwidth]{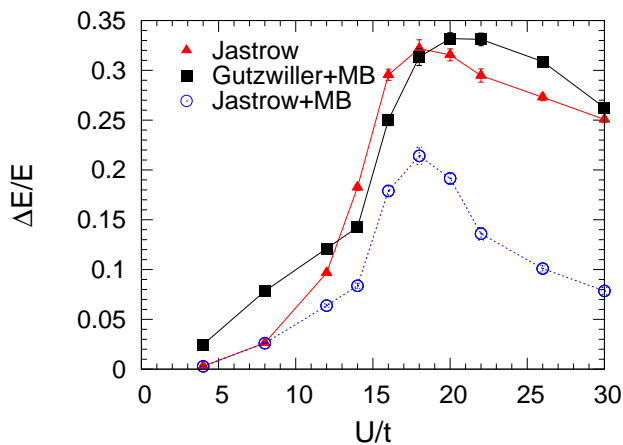}
\caption{\label{accuracy3d}
Accuracy of different variational wave functions as a function of $U/t$ for
the $10 \times 10\times 10$ cluster and $1000$ bosons. The symbols are the 
same as in Fig.~\ref{fig:accuracy1d}.}
\end{figure}

\begin{figure}
\includegraphics[width=\columnwidth]{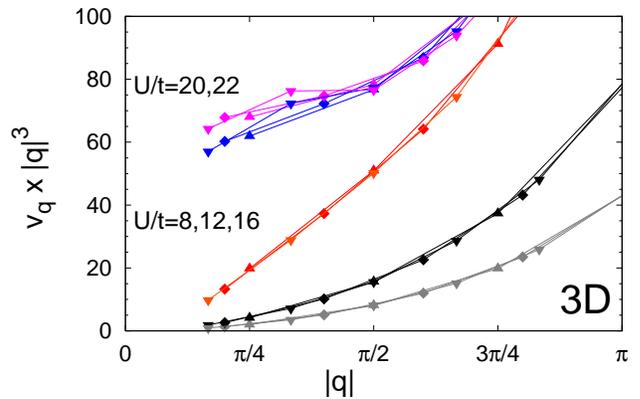}
\caption{\label{jas3d}
Jastrow parameters $v_q$ multiplied by $|q|^3$ as a function of $|q|$ 
for $8 \times 8 \times 8$ (circles), $10 \times 10 \times 10$ (squares), and 
$12 \times 12 \times 12$ (triangles) clusters [along the (1,0,0) direction].}
\end{figure}

\begin{figure}
\includegraphics[width=\columnwidth]{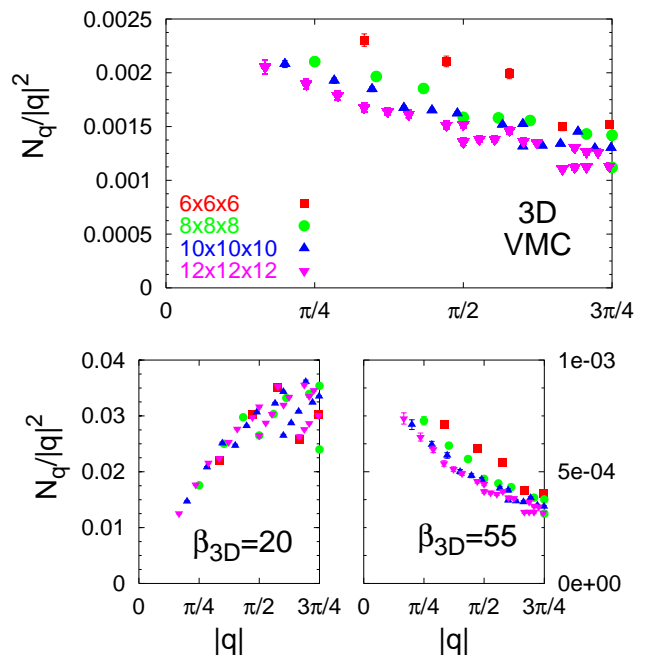}
\caption{\label{nq3d}
Upper panel: Variational results for the density structure factor $N_q$ 
for 3D and $U/t=20$. Lower panels: $N_q$ for non-optimized wave functions with
$v_q \sim \beta_{3D}/|q|^3$ for two values of $\beta_{3D}$.}
\end{figure}

Let us move now to the 3D case. Here, we mostly restrict our analysis to the 
variational method that allows us to assess rather large sizes.
We start by considering the accuracy of the different variational states as a 
function of the interaction $U/t$, see Fig.~\ref{accuracy3d}. 
It turns out that in the large-$U$ region the most accurate state contains 
both the long-range Jastrow term and the many-body term, as it occurs in 2D. 
Moreover, in analogy to 1D and 2D, the presence of a phase transition upon 
increasing $U/t$ is clearly signaled by the sudden change in the small-$q$ 
behavior of the Jastrow factor, see Fig.~\ref{jas3d}. Here, its behavior 
changes drastically from $v_q \sim \alpha/|q|$ to a more diverging 
$v_q\sim 1/|q|^3$ in the insulating regime. In particular, the sudden change 
of behavior allows us to locate the transition around $U_c/t\simeq 18$, which 
is close to the critical value of recent Monte Carlo 
simulations.~\cite{prokofev} 

Focusing on the large-$U$ region of the phase diagram, let us discuss the 
implications of the singular $v_q\sim 1/|q|^3$ Jastrow potential. First of all,
following the same arguments of Ref.~\onlinecite{reatto}, we can assert that
the strong singular character of the Jastrow potential is able to empty the 
condensate, whereas a less singular $v_q \sim 1/q^2$ would not lead to 
${\tt n}_0 \to 0$ in the thermodynamic limit. Remarkably, even though 
$v_q\sim 1/|q|^3$, the structure factor in the insulator has the correct 
$N_q\sim q^2$ behavior, see Fig.~\ref{nq3d}. In turn, this implies that 
Eq.~(\ref{nqvsvq}) does not hold. In order to prove 
more firmly that $v_q\sim \beta_{3D}/|q|^3$ can indeed lead to $N_q\sim q^2$, 
we have calculated the density structure factor with a {\it non-optimized} 
wave function of the form~(\ref{wfj}) with $v_q\sim \beta_{3D}/|q|^3$ and for 
different values of $\beta_{3D}$. As shown in Fig.~\ref{nq3d}, for small 
$\beta_{3D}$ we obtain $N_q\sim |q|^3$, implying that Eq.~(\ref{nqvsvq}) is 
qualitatively correct in this case. However, above a critical $\beta^*_{3D}$, 
the small-$q$ behavior of the density structure factor turns into 
$N_q \sim q^2$, signaling a remarkable breakdown of Eq.~(\ref{nqvsvq}).
From Fig.~\ref{jas3d} it turns out that the optimal value of 
$\beta_{3D}$ that we get variationally at the superfluid-insulator transition 
is larger than $\beta^*_{3D}$, confirming that $N_q \sim q^2$ in the insulating
phase. Most importantly, the change of behavior as a function of $\beta_{3D}$ 
is consistent with the binding-unbinding phase-transition recently uncovered 
in a classical 3D gas with a $1/|q|^3$ potential.~\cite{3dcgm}
Once again, as in 1D and 2D, the Mott insulating wave function in 3D is found 
to be closely related to the low-temperature confined phase of a classical 
model, where opposite charges tightly bound. 

\section{2D systems with long-range interaction}\label{reslr}

In the previous paragraphs, we have shown that, in the Hubbard 
model~(\ref{hambose}), the superfluid regime can be described by a long-range 
Jastrow wave function with $v_q \sim \alpha/|q|$. By increasing the on-site 
interaction $U$, our variational approach describes a continuous transition to 
an insulating phase that corresponds to the confined phase of a classical model 
of interacting particles with opposite charge. In particular, we have found 
evidences that the 2D Mott insulating wave function corresponds to a classical 
model with $-\log(q)/q^2$ potential rather than to a 2D CGM with potential 
$\beta/q^2$ in the confined phase ($\beta>\beta^*\geq 4\pi$).  
This result, as we discussed, has a physical importance since the confined 2D 
CGM has power-law correlations, not possible in a Mott insulating phase. 
Nevertheless, it would be interesting to search for bosonic Hamiltonians whose 
ground state can be described by the variational wave function~(\ref{wfjmb}) 
with a Jastrow potential $v_q \sim \beta/q^2$.  

\begin{figure}
\includegraphics[width=\columnwidth]{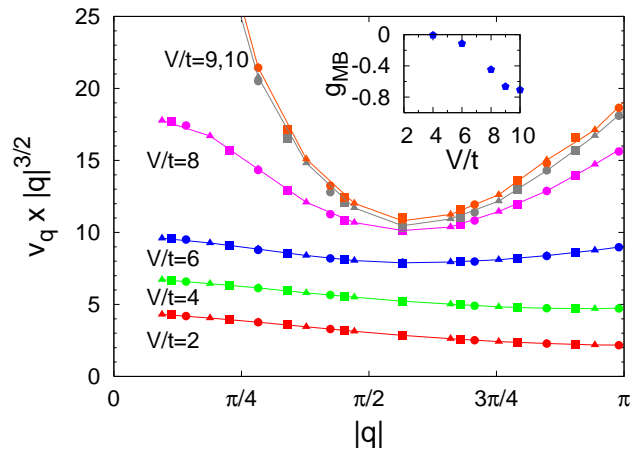}
\caption{\label{fig:jaspois}
Jastrow parameters $v_q$ multiplied by $|q|^{3/2}$ as a function of $|q|$
for the potential of Eq.~(\ref{Upoisson}) with different $V/t$ and 
$20 \times 20$ (circles), $26 \times 26$ (squares), and $30 \times 30$ 
(triangles) clusters. Inset: The many-body variational parameter $g_{MB}$ as a 
function of $V/t$.}
\end{figure}

\begin{figure}
\includegraphics[width=\columnwidth]{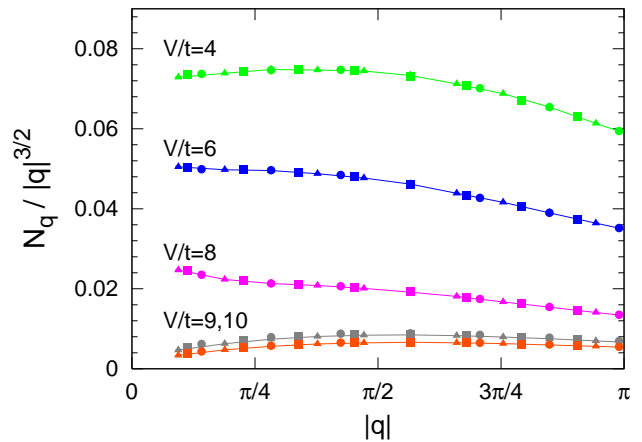}
\caption{\label{fig:nqpois}
Variational results for the density structure factor $N(q)$ divided by
$|q|^{3/2}$ for the potential of Eq.~(\ref{Upoisson}) with different $V/t$ 
and $20 \times 20$ (circles), $26 \times 26$ (squares), and $30 \times 30$ 
(triangles) clusters.}
\end{figure}

\begin{figure}
\includegraphics[width=\columnwidth]{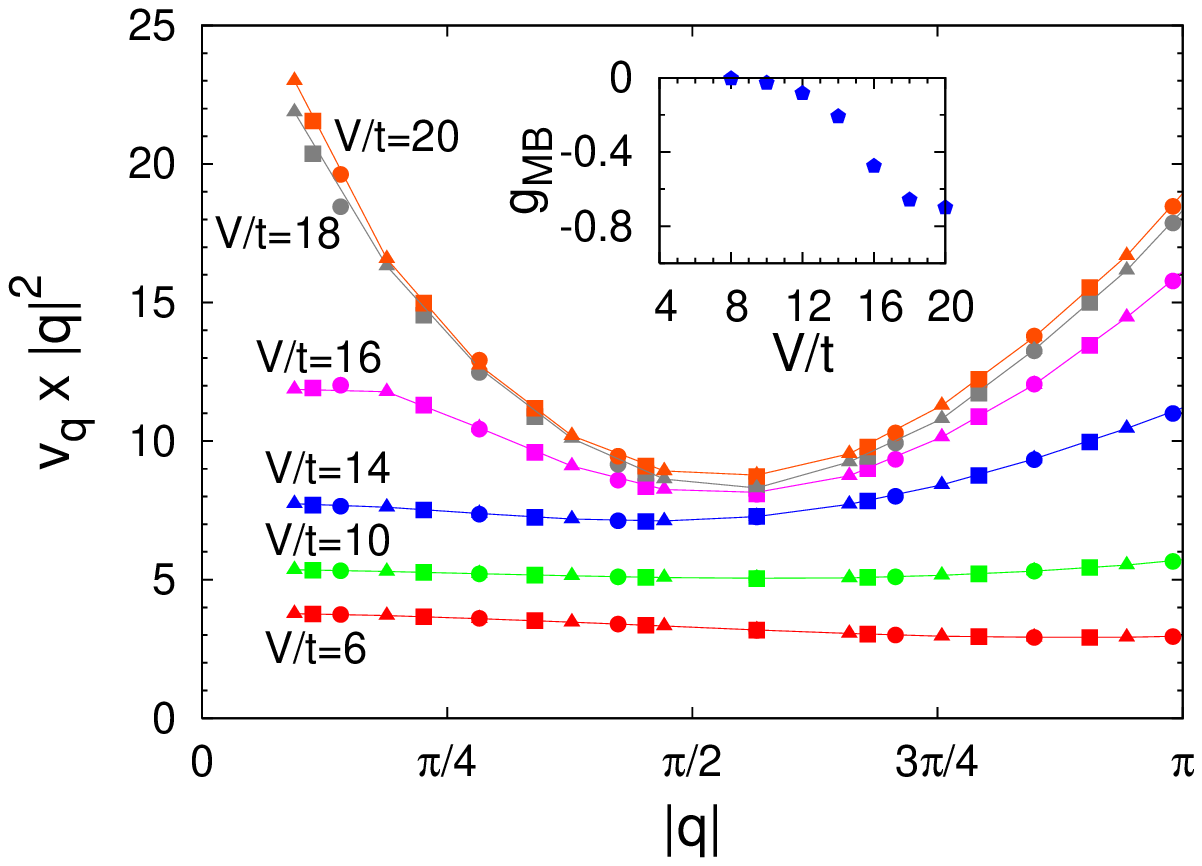}
\caption{\label{fig:jasUq2}
The same as in Fig.~\ref{fig:jaspois}, for the potential of Eq.~(\ref{Uq2}).}
\end{figure}

\begin{figure}
\includegraphics[width=\columnwidth]{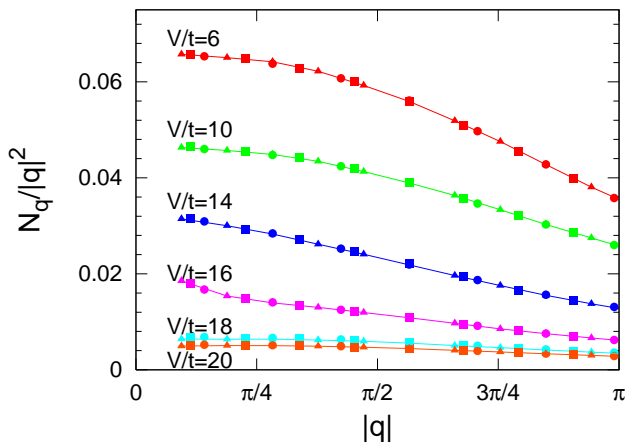}
\caption{\label{fig:nqUq2}
The same as in Fig.~\ref{fig:nqpois}, for the potential of Eq.~(\ref{Uq2}).}
\end{figure}

For that purpose, we consider the more general Hubbard model of 
Eq.~(\ref{hamboselr}) with a long-range interaction $\Omega(r)$. Let us start 
by considering the realistic case of a Coulomb potential, namely 
$\Omega(r) \sim 1/r$ that in 2D leads to Eq.~(\ref{Upoisson}). Then, we can  
vary its strength $V$ to drive the system across a superfluid-insulator
transition. The small-$q$ behavior of the optimized Jastrow parameters is 
shown in Fig.~\ref{fig:jaspois}. For $V/t \lesssim 8$, we obtain that 
$v_q \sim 1/|q|^{3/2}$. On the contrary, for larger values of the interactions,
$v_q$ becomes more singular and in particular is best fitted by 
$-\log(q)/q^2$, just like what we found for short-range interaction.  
The structure factor in the weak-coupling phase behaves as 
$N_q \sim |q|^{3/2}$, see Fig.~\ref{fig:nqpois}, compatible with a superfuid 
phase with 2D plasmons. We mention that a similar behavior has been found in 
continuum models at high densities both analytically~\cite{apaja} and 
numerically.~\cite{depalo} In the strong-coupling regime, the insulating 
behavior $N_q \sim q^{2}$ is recovered. These results are confirmed by GFMC 
(not shown), though the critical $V/t$ is slightly decreased, i.e., 
$V_c/t \sim 7$. In this case, as before with short-range interaction, the 
optimized Jastrow never behaves as the potential of a CGM.
 
Therefore, let us turn to the more singular interaction given by 
Eq.~(\ref{Uq2}), leading to $\Omega(r) \sim -\log(r)$. In this case the 
potential in $q$-space is given by $\Omega(q) \sim 1/q^2$ and we expect, 
similarly to what happens in the continuum for high density,~\cite{magro} 
that in the weak-coupling regime also $v_q \sim \beta/q^2$. Indeed, as shown 
in Fig.~\ref{fig:jasUq2}, this is the case for $V/t\lesssim 16$. Above this 
value, $v_q$ becomes once again of the form $-\log(q)/q^2$, just like in all 
previous examples. We note that the values of $\beta$ extracted in the weak 
coupling phase seem to be all smaller than $\beta^*$ of the 2D 
Berezinskii-Kosterlitz-Thouless phase transition, although we can not exclude 
by the numerical data that $\beta \to \beta^*$ at the transition. 
This indicates that the CGM that corresponds to the variational wave function 
is in the plasma phase, with exponential decaying density-density correlation 
functions. Indeed, the structure factor behaves like $N_q \sim q^{2}$ for all 
the coupling strengths $V/t$, see Fig.~\ref{fig:nqUq2}. We believe that the 
weak-coupling phase has to be identified with the algebraic long-range ordered 
phase found at high density in the continuum model by Magro and 
Ceperley.~\cite{magro} This phase is characterized by absence of condensate 
but by a power-law decay of the single-particle density matrix. 
On the contrary, the strong-coupling phase with Jastrow potential 
$v_q\sim -\log(q)/q^2$ must correspond to a genuine Mott insulator with all 
correlation functions decaying exponentially. 

\section{Conclusions}\label{conc}

We have shown that the long-range Jastrow wave function gives a consistent 
picture of the superfluid-insulator transition of the bosonic Hubbard model 
in all spatial dimensions and also in the presence of long-range interaction. 

In one dimension the variational results are compatible with a 
Berezinskii-Kosterlitz-Thouless phase transition between the quasi-long-range 
ordered gapless phase and the Mott insulator. From the point of view of the 
variational wave function, the gapless phase is characterized by a Jastrow 
potential $v_q \sim \alpha/|q|$, while the insulating one by $v_q\sim 1/q^2$.

In two dimensions we have evidences of a second-order phase transition between 
a superfluid phase and a Mott insulator. Here, the superfluid wave function 
still has $v_q \sim \alpha/|q|$, compatible with the existence of sound modes, 
while the insulating wave function is characterized by a more singular 
$v_q\sim -\log(q)/q^2$. This singular behavior in the Mott phase does not 
change even if long-range interaction is considered. For instance, for a 
Coulomb interaction $\Omega(r) \sim 1/r$, the Jastrow potential in the 
superfluid phase changes into $v_q\sim 1/|q|^{3/2}$, compatible with the 
existence of 2D plasmons, yet the insulating wave function has still 
$v_q\sim -\log(q)/q^2$. For an interaction $\Omega(r) \sim \log(r)$, we observe
a transition between an algebraic long-range ordered phase,~\cite{magro} 
characterized by $v_q\sim 1/q^2$, and a Mott insulating phase, once again with 
$v_q\sim -\log(q)/q^2$. 

In three dimensions, the Mott transition as revealed by the behavior of the 
Jastrow potential becomes much more evident. As usual, the superfluid phase 
has $v_q \sim \alpha/|q|$. In this case, however, the Mott insulating wave 
function has a much more singular $v_q\sim 1/|q|^3$, still with a structure 
factor that correctly behaves as $N_q\sim q^2$.

This work was mainly focused on bosons, but a similar variational wave function 
can be applied also to fermionic models, with the same key ingredient, i.e., 
a long-range Jastrow factor that drives the metal-insulator 
transition, in spite of the uncorrelated wave function being 
metallic.~\cite{capello,capello3} According to this picture, the Mott 
insulating state should be produced by a sufficiently long-range 
Jastrow potential able to bind opposite-charge fluctuations. Based on the 
bosonic results, we may argue that a Jastrow potential $v_q \sim 1/|q|^D$ 
may work even for fermions in any dimensions $D>2$, $D=2$ playing somehow the 
role of the lowest critical dimension, $v_q\sim -\log(q)/q^2$. 
As the interaction strength is decreased, an unbinding transition must takes 
place in the variational wave function,  turning the Mott insulator into a 
correlated metal and providing a simple physical picture of the long-standing 
but still actual and very attractive phenomena which is the Mott transition.   

We thank important discussions with F. Alet and S. de Palo.
This work has been partially supported by CNR-INFM and COFIN 2004 and 2005.

\appendix 
\section{Details for the calculation of the stiffness}

In this Appendix, we give some details for the zero-temperature GFMC 
calculation of the stiffness $D_s$. In this respect, we consider the
standard Peierls substitution and introduce an electromagnetic field in the 
Hamiltonian by replacing the hopping term between two sites $R_i$ and $R_j$ 
by a suitable complex hopping:~\cite{peierls}
\begin{equation}
t \to t \exp \left \{i \int_{R_i}^{R_j} dR \; A(R) \right \},
\end{equation}
where we can consider $A(R)=(A_x,0)$. Then, the first derivative of the
energy $E[A]$ with respect to this field gives the charge current, containing 
both the paramagnetic and diamagnetic contributions. This current must vanish
when $A_x \to 0$, since the Hamiltonian is real. For a bosonic system, the 
second derivative of the energy $E[A]$ represents the charge 
stiffness.~\cite{kohn}

Within GFMC it is possible to compute the ground-state energy $E[A]$ for 
arbitrary time-independent static field $A_x$. Let us denote by ${\cal H}_A$ 
and ${\cal H}$ the Hamiltonian in presence and absence of the field, 
respectively and consider: 
\begin{equation}
Z_\tau[A] = \frac{1}{\tau} 
\frac{\langle \Phi | e^{-\tau {\cal H}_A} |\Psi_0 \rangle}
{\langle \Phi | e^{-\tau {\cal H}} |\Psi_0 \rangle},
\end{equation} 
where $|\Psi_0 \rangle$ is the ground state of ${\cal H}$, with $E_0$ energy, 
and $|\Phi \rangle$ is the guiding wave function of the GFMC method.
In the large-$\tau$ limit, we have that
\begin{equation}
Z_\tau[A] \sim \frac{1}{\tau} 
\frac{\langle \Phi | \Psi_0^A \rangle \langle \Psi_0^A | \Psi_0 \rangle} 
{\langle \Phi | \Psi_0 \rangle} \; e^{-\tau (E_0^A-E_0)},
\end{equation} 
where $|\Psi_0^A \rangle$ and $E_0^A$ are the ground state eigenfunction and
eigenvalue of ${\cal H}_A$, respectively. By taking the second derivative of 
$Z_\tau[A]$ with respect to $A$ and then considering $A=0$, we obtain the 
charge stiffness (up to $1/\tau$ corrections).

This quantity can be obtained by sampling statistically the unperturbed Green's
function $G_{x^\prime,x}= -\Phi_{x^\prime} {\cal H}_{x^\prime,x}/\Phi_x = 
p_{x^\prime,x} b_x$, where $p_{x^\prime,x}$ is a stochastic matrix that defines
the Markov chain and $b_x$ is a normalization factor. In this way, the 
walker $|x\rangle$ is distributed according to the variational distribution 
$|\langle x|\Phi \rangle|^2$ and, in order to obtain the true ground state,
the weight $G^\tau$ must be considered. Then
\begin{equation}
Z_\tau[A] = \frac{1}{\tau} \frac{\sum_n G^\tau_n[A]}{\sum_n G^\tau_n}   
\end{equation}
where the index $n$ indicates the Markov chain iteration, that is defined by
the transition probability $p_{x^\prime,x}$, and
\begin{eqnarray}
G^\tau_n &=& \exp \left \{ \int_0^\tau d\tau^\prime e_L[x(\tau^\prime)] \right \} \\
G^\tau_n[A] &=& G^\tau_n \exp \left \{ i \int_{R_{x(0)}}^{R_{x(\tau)}} dR \; A(R) \right \}
\end{eqnarray}
are the correcting factors of the GFMC method; $e_L(x)$ indicates the local
energy and $R_{x}$ indicates the site where the particle moves within the
GFMC algorithm. By considering the second derivative of $Z_\tau[A]$ with 
respect to $A_x$, we obtain Eq.~(\ref{stiffness}). As usual, many walkers can 
be considered with the branching technique in order to reduce the variance of 
the correcting factors $G^\tau_n$.

\end{document}